\documentclass[submission,copyright,creativecommons]{eptcs}
\providecommand{\event}{FLACOS 2012} 
\usepackage{breakurl}             
\usepackage[english]{babel}
\usepackage[latin1]{inputenc}
\usepackage{multicol}

\newcommand{\codiag}{\textit{C-O Diagrams\,}}

\usepackage{latexsym,makeidx,graphicx,color,colortbl}
\usepackage{moreverb,amsmath,amssymb,dsfont,stmaryrd}
\usepackage{pstricks,pst-node}
\usepackage{paralist}
\usepackage{array}
\usepackage{hyperref}
\usepackage[latin1]{inputenc}
\usepackage{subfig}

\newtheorem{definition}{Definition}
\newtheorem{example}{Example}

\newcommand{\bdfn}{\begin{definition} \begin{rm}}
\newcommand{\edfn}{\vspace{-2ex}
{\flushright $\Box$\\
\mbox{}\vspace{-2ex}
} \end{rm} \end{definition}}
\newcommand{\edfnt}{ \end{rm} \end{definition}}
\newcommand{\bthm}{\begin{theorem} \begin{rm}}
\newcommand{\ethm}{\vspace{-4ex}{\flushright $\Box$\\
\mbox{}\vspace{-4ex}} \end{rm} \end{theorem}}
\newcommand{\bprop}{\begin{proposition} \begin{rm}}
\newcommand{\eprop}{\vspace{-4ex}{\flushright $\Box$\\
\mbox{}\vspace{-4ex}} \end{rm} \end{proposition}}
\newcommand{\bcor}{\begin{corollary}\begin{rm}}
\newcommand{\ecor}{\vspace{-4ex}{\flushright $\Box$\\
\mbox{}\vspace{-4ex}} \end{rm} \end{corollary}}
\newcommand{\blem}{\begin{lemma} \begin{rm}}
\newcommand{\elem}{\vspace{-4ex}{\flushright $\Box$\\
\mbox{}\vspace{-4ex}} \end{rm} \end{lemma}}
\newcommand{\bfact}{\begin{fact} \begin{rm}}
\newcommand{\efact}{\vspace{-4ex}{\flushright $\Box$\\
\mbox{}\vspace{-4ex}} \end{rm} \end{fact}}
\newcommand{\bex}{\begin{example} \begin{rm}}
\newcommand{\eex}{\vspace{-2ex}{\flushright $\Box$\\
\mbox{}\vspace{-2ex}} \end{rm} \end{example}}

\newcommand{\real}{\rm I\! R^+}
\newcommand{\act}{\ensuremath{\mathit{ACT}}}

\long\def\comment#1{}

\newcommand{\comentario}[1]{}

\newbox\arriba
\newbox\abajo
\newbox\CaracterInterno
\newbox\CaracterDerecha
\newdimen\anchura
\def\MacrosTranGeneral#1#2#3#4#5#6{%
  \setbox\CaracterInterno=\hbox{\mathsurround=0pt$\mathord#4$}
  \setbox\CaracterDerecha=\hbox{\mathsurround=0pt$\mathord#3$}
  \setbox\arriba=\hbox{$#1#2$}
  \setbox\abajo=\hbox{\mathsurround=0pt%
                      \anchura=\wd\arriba%
                      \advance \anchura by 0.5em%
                      \divide \anchura by \wd\CaracterInterno%
                      \multiply \anchura by \wd\CaracterInterno%
                      \copy\CaracterInterno\kern\SeparacionInternaFlecha
                      \hbox to \anchura{%
                          $\cleaders%
                            \hbox{\kern\SeparacionInternaFlecha\copy\CaracterInterno}
                            \hfill$}%
                      \kern\SeparacionExternaFlecha\copy\CaracterDerecha}
  \mathrel{{\buildrel\vbox{\copy\arriba \kern\SeparacionFlechaArriba} %
    \over{\copy\abajo^{#6}}}_{#5}}
  }
\def\MacrosTranGeneralProp#1#2#3#4#5{\mathchoice%
  {\MacrosTranGeneral{\scriptstyle}{#1}{#2}{#3}{#4}{#5}}
  {\MacrosTranGeneral{\scriptstyle}{#1}{#2}{#3}{#4}{#5}}
  {\MacrosTranGeneral{\scriptscriptstyle}{#1}{#2}{#3}{#4}{#5}}
  {\MacrosTranGeneral{\scriptscriptstyle}{#1}{#2}{#3}{#4}{#5}}}

\def\MacrosTran#1{%
  \def\SeparacionInternaFlecha{-0.3em}
  \def\SeparacionExternaFlecha{-0.5em}
  \def\SeparacionFlechaArriba{-3pt}
  \MacrosTranGeneralProp{#1}{\rightarrow}{-}{}{}}
\def\MacrosNoTran#1{%
  \def\SeparacionInternaFlecha{-0.3em}
  \def\SeparacionExternaFlecha{-0.5em}
  \def\SeparacionFlechaArriba{-3pt}
  \MacrosTranGeneralProp{#1\kern 0.5em}{{\not\rightarrow}}{-}{}{}}
\def\MacrosVTran#1{%
  \def\SeparacionInternaFlecha{-0.2em}
  \def\SeparacionExternaFlecha{-0.5em}
  \def\SeparacionFlechaArriba{0pt}
  \MacrosTranGeneralProp{#1}{\Rightarrow}{=}{}{}}

\def\tran#1{\ensuremath\mathbin{\MacrosTran{#1}}}
\def\vtran#1{\ensuremath\mathbin{\MacrosVTran{#1}}}

\newcommand{\traces}{\ensuremath{\mathsf{tr}}}

\newcommand{\permission}{P}
\newcommand{\tviolation}{\mathsf{V}}
\newcommand{\tsatis}{\mathsf{S}}
\newcommand{\tpermission}{\mathsf{P}}
\newcommand{\clean}{\ensuremath{\mathsf{clean}}}
\newcommand{\good}{\ensuremath{\mathsf{good}}}
\newcommand{\hide}[1]{\ensuremath{\mathsf{hide}_{#1}}}
\newcommand{\conf}{\mathbin{\mathsf{conf}}}

\renewcommand{\emptyset}{\varnothing}
\newcommand{\leqp}{\mathbin{\leq_P}}

\title{Conformance Verification of Normative Specifications using C-O Diagrams}

\author{Gregorio D\'{i}az$^1$, Luis Llana$^2$, Valent\'{i}n Valero$^1$ and Jose A. Mateo$^1$
\institute{1. Computer Science Dept. University of Castilla-La Mancha}
\email{[gregorio.diaz,valentin.valero,joseantonio.mateo]@uclm.es}
\institute{2. Computer Science Dept. Complutensis University of Madrid}
\email{llana@fdi.ucm.es}
}

\def\titlerunning{Conformance Verification of C-O Diagrams}
\def\authorrunning{G. D\'{i}az, L. Llana, V. Valero, \& J.A. Mateo}

\begin{document}

\maketitle

\begin{abstract}

C-O Diagrams have been introduced as a means to have a visual representation of normative texts and electronic contracts, where it is possible to represent the obligations, permissions and prohibitions of the different signatories, as well as what are the penalties in case of not fulfillment of their obligations and prohibitions. In such diagrams we are also able to represent absolute and relative timing constrains.
 In this paper we consider a formal semantics for C-O Diagrams based on a network of timed automata and we present several relations to check the consistency of a contract in terms of realizability, to analyze whether an implementation satisfies the requirements defined on its contract, and to compare several implementations using the executed permissions as criteria.

\end{abstract}



\section{Introduction}

In the software context, the term {\it contract} has traditionally been used as a metaphor to represent limited kinds of ``agreements'' between software elements at different levels of abstraction. The first use of the term in connection with software programming and design  was done by Meyer in the context of the language Eiffel ({\it programming-by-contracts}, or {\it design-by-contract}). 
This notion of contracts basically relies on the Hoare notion of pre and post-conditions and invariants. Though this paradigm has proved to be useful for developing object oriented systems, it seems to have shortcomings for novel development paradigms such as service-oriented computing and component-based development. These new applications have a more involved interaction and therefore require a more sophisticated notion of contracts. As a response, behavioural interfaces have been proposed to capture richer properties than simple pre and post-conditions \cite{Hatcliff2009}. Here it is possible to express contracts on the history of events, including causality properties. 
In the context of SOA, there are different service contract specification languages, like ebXML, WSLA, and WS-Agreement.
These standards and specification languages suffer from one or more of the following problems: They are restricted to bilateral contracts, lack of formal semantics (so it is difficult to reason about them), their treatment of functional behaviour is rather limited and the sub-languages used to specify, for instance, security constraints are usually limited to small application-specific domains. The lack of suitable languages for contracts in the context of SOA is a clear conclusion of the survey \cite{OR08csc} where a taxonomy is presented.

\comment{
More recently, some researchers have investigated how to adapt deontic logic \cite{McNamara2006} to define (consistent) contracts targeted to software systems where the focus is on the normative notions of obligation, permission and prohibition, including sometimes exceptional cases (e.g., \cite{PS09cl}).
Independently of the application domain, there still is need to better fill the gap between a contract understood by non-experts in formal methods (for its use), its logical representation (for reasoning), and its internal machine-representation  (for runtime monitoring, and to be manipulated by programmers). We see two possible ways to bridge this gap: i) to develop suitable techniques to get a good translation from contracts written in natural language into formal languages, and ii) to provide a graphical representation (and tools) to manipulate contracts at a high level, with formal semantics supporting automatic translation into the formal language.
We take in this paper the second approach.
}

In \cite{MCD+10} \codiag  were introduced, a graphical representation not only for electronic contracts but also for the specification of any kind of normative text (Web service composition behaviour, software product lines engineering, requirements engineering, \ldots). \codiag\ allow the representation of complex clauses describing the obligations, permissions, and prohibitions of different signatories (as defined in deontic logic \cite{McNamara2006}), as well as {\it reparations} describing contractual clauses in case of not fulfillment of obligations and prohibitions. Besides, \codiag\ permit to define real-time constraints. In \cite{Martinez2011} some of the satisfaction rules needed to check if a timed automaton satisfies a \textit{C-O Diagram} specification were defined. In \cite{MCDS2012}, \codiag\ are equipped with a formal semantics based on a transformation of these diagrams into a network of timed automata (NTA). 
The contribution of this work pursues the further development of our
previous work. This time we will focus on the development of different
relations to check the consistency of contracts, to seek whether an
implementation conforms a given contract and to compare several implementations.
To achieve this goal, we consider a semantics in terms of NTAs and we establish
relations with the implementations also written in terms of NTAs.



\section{Related Work}
\label{Related}

The use of deontic logic for reasoning about contracts is widely spread in the literature since it was proposed in \cite{Dignum1995} for modelling communication processes. In \cite{Marjanovic2001} Marjanovic and Milosevic present their initial ideas for formal modelling of e-contracts based on deontic constraints and verification of deontic consistency, including temporal constraints. In \cite{Governatori2006} Governatori et al. go a step further providing a mechanism to check whether business processes are compliant with business contracts. They introduce the logic FCL to reason about the contracts, based again on deontic logic. In \cite{Lomuscio2008} Lomuscio et al. provides another methodology to check whether service compositions are compliant with e-contracts, using WS-BPEL to specify both, all the possible behaviours of each service and the contractually correct behaviours, translating these specifications into automata supported by the MCMAS model checker to verify the behaviours automatically.

\comment{
The approach followed by \textit{C-O Diagrams} is inspired by the formal language $\cal CL$ \cite{PS09cl}. In this language a contract is also expressed as a composition of obligations, permissions and prohibitions over actions, and the way of specifying reparations is the same that in our model. However, $\cal CL$ does not support neither the specification of agents nor timing constraints natively, so they have to be encoded in the definition of the actions. In \cite{Solaiman2003} Solaiman et al. show how relevant parts of contracts can be described by means of Finite State Machines (FSMs), using these FSMs to check the correctness of the contract specification, detecting any undesirable ambiguity, whereas our approach seeks to check if the contract specified satisfies some properties of interest by using the verifier of the UPPAAL tool. In \cite{Desai2008} Desai et al. also automate reasoning about the correctness the contract specification, but in this case representing contracts formally as a set of commitments.
}

None of the previous works provides a visual model for the definition of contracts.
However, there are several works that define a meta-model for the specification of e-contracts which purpose is their enactment or enforcement. In \cite{Chiu2003} Chiu et al. present a meta-model for e-contract templates written in UML, where a template consists of a set of contract clauses of three different types: obligations, permissions and prohibitions. These clauses are later mapped into ECA rules for contract enforcement purposes,   but the templates do not include any kind of reparation or recovery associated to the clauses. In \cite{Krishna2005} Krishna et al. another meta-model based on entity-relationship diagrams is proposed to generate workflows supporting e-contract enactment. This meta-model includes clauses, activities, parties and the possibility of specifying exceptional behaviour, but this approach is not based on deontic logic and says nothing about including real-time aspects natively. 


\section{C-O Diagrams Syntax and Semantics}
\label{Model}

\begin{figure}

\begin{center}
  \includegraphics[width=14cm]{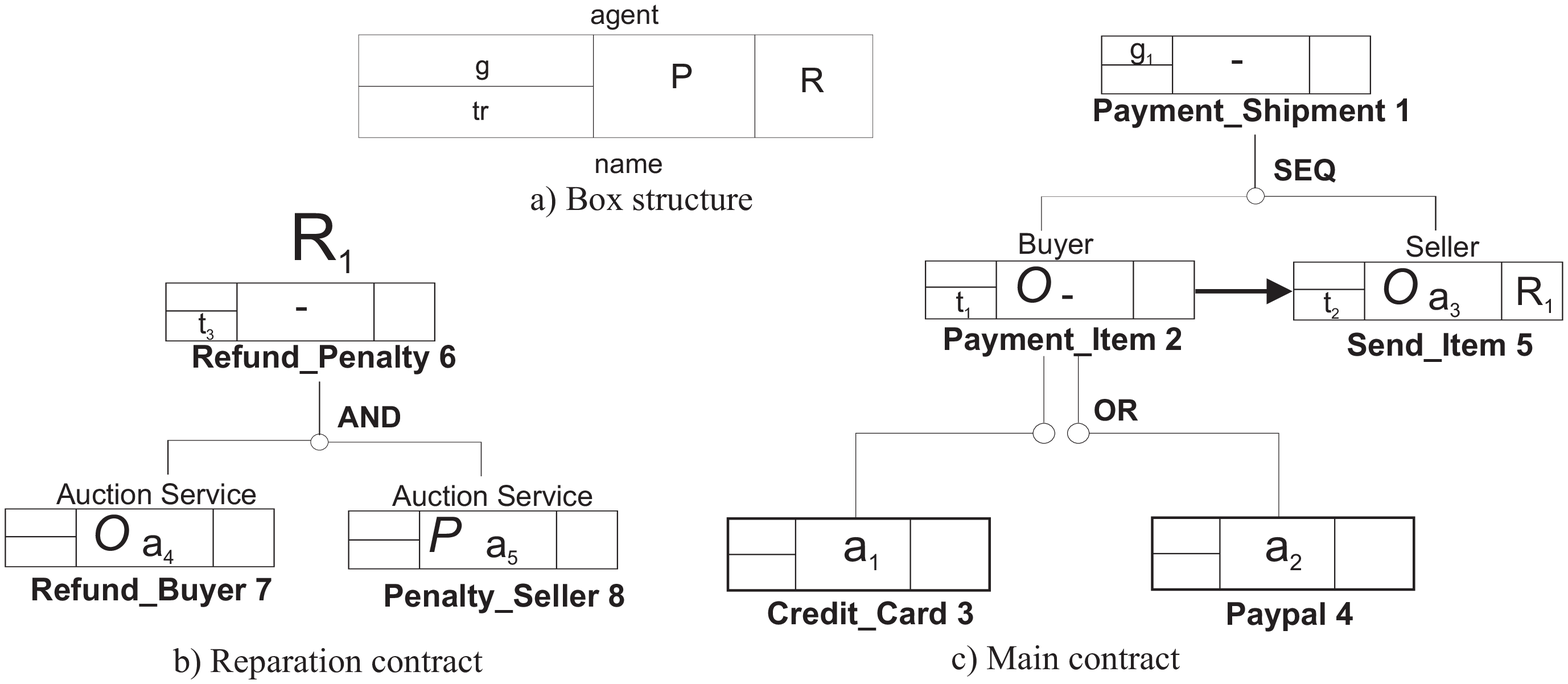}
\caption{\codiag examples}
\label{3figs}
\end{center}
\vspace{-1.0cm}
\end{figure}

We first introduce a motivation example to understand the
diagrams in an easy way. Figure \ref{3figs} consists of three sub-figures, a) depicting a
basic structure of a clause, and, b) and c) depicting our running example.
This example consists in the payment and shipment of an item previously sold during an online auction.
Thus the action starts after  the auction has finished, that is,  if the bid placed by the \textit{buyer} is the highest one, then the activities concerning the payment and the shipment of the item start. First, the \textit{buyer} has \textbf{three days} to perform the payment, which can be done by means of credit card or PayPal. After the payment has been performed, the \textit{seller} has \textbf{fourteen days} to send the item to the \textit{buyer}. If the item is not received within this period of time, the \textit{auction service} has \textbf{seven days} to refund the payment to the \textit{buyer} and can penalize the \textit{seller} in some way.


At first sight,  the figures are top down hierarchical structures with several boxes and branches. In Figure \ref{3figs}, we can observe several examples.
 At the top-left hand side of this figure, Figure \ref{3figs}-a, we can observe the basic construction element called \textbf{box}, also referred as proposition or clause. It is divided into four fields. The \textit{guard} \textbf{g} specifies the conditions under which the contract clause must be taken into account (boolean expression). The \textit{time restriction} \textbf{tr} specifies the time frame during which the contract clause must be satisfied (deadlines, timeouts, etc.). The \textit{propositional content} \textbf{P}, on the centre, is the main field of the box, and it is used to specify normative aspects such as obligations (\textbf{O}), permissions (\textbf{P}) and prohibitions (\textbf{F}), that are applied over actions, and/or the specification of the actions themselves. The last field of these boxes, on the right-hand side, is the \textit{reparation} \textbf{R}. This reparation, if specified by the contract clause, is a reference to another contract that must be satisfied in case the main norm is not satisfied (a \textit{prohibition} is violated or an \textit{obligation} is not fulfilled, there is no reparation for \textit{permission}), considering the clause eventually satisfied if this reparation is satisfied. Each box has also a name at the bottom part and an agent at the top part. 

These are the basic boxes, which can be composed by using some refinements. Refinements are classified into three types: joining \textit{AND-refinements}, disjunctive \textit{OR-refinements} and sequential \textit{SEQ-refinement}. Joining refinements require that all the hanging propositions should be accomplished to declare the upper proposition accomplished; on the contrary, disjunctive propositions only require one to be accomplished; whereas, sequential propositions require a left-to-right ordered sequential satisfaction of every proposition to obtain the same result. In Figure \ref{3figs}-c, the root box, which only shows the name and guard $g_1$ (this guard checks if the buyer is the auction winner) is decomposed into two sub-clauses via sequential composition, that is, first the one on the left hand side, \textit{Payment\_Item} and, afterwards, the one on the right hand side, \textit{Send\_Item}. The first one is the \textbf{obligation} (\textit{O}) of payment with the temporal restriction $t_1$, three days in this case, then this obligation is decomposed via an \textit{OR-refinement} into \textbf{Clause 3} and \textbf{Clause 4} composing the actions of paying by credit card or PayPal by means of an \textit{OR-refinement}. On the right-hand side we have the \textbf{obligation} (\textit{O}) specified in \textbf{Clause 5}, which has been called \textit{Send\_Item}, including the real-time constraint $t_2$ 14 days and a reference to reparation $R_1$.

Since reparations are references to new contracts, in Figure \ref{3figs}-b we can see the diagram corresponding to reparation $R_1$. It has been called \textit{Refund\_Penalty}, including the real-time constraint $t_3$, and it is decomposed into two subclauses by means of an \textit{AND-refinement}. The subclause on the left corresponds to the \textbf{obligation} (\textit{O}) specified in \textbf{Clause 7}, which has been called \textit{Refund\_Buyer}, and the subclause on the right corresponds to the \textbf{permission} (\textit{P}) specified in \textbf{Clause 8}, which has been called \textit{Penalty\_Seller} regarding the possibility of performing some kind of penalization over the seller by the \textit{Auction Service}.



The \textbf{syntax} of \codiag was first presented in \cite{MCD+10}. Next, we just present a brief description of the EBNF grammar followed in the diagrams:


{
\small

\begin{center}
\begin{tabular}{rclp{0.25cm}|p{0.25cm}rcl}
    $C$ & $:=$ &  $(agent,name,g,tr,O(C_2),R) \,|$ & & & $C_1$ & $:=$ &  $C \,(And \; C)^+\,|\, C \,(Or \; C)^+\,|\, C \,(Seq \; C)^+$ \\

    &  &  $(agent,name,g,tr,P(C_2),\epsilon) \,|$  & & & $C_2$ & $:=$ &  $a \,|\, C_3 \,(And \; C_3)^+\,|\, C_3 \,(Or \; C_3)^+\,|\, C_3 \,(Seq \; C_3)^+$ \\

    &  &  $(agent,name,g,tr,F(C_2),R) \,|$ & & & $C_3$ & $:=$ & $(\epsilon,name,\epsilon,\epsilon,C_2,\epsilon)$ \\

    &  &  $(\epsilon,name,g,tr,C_1,\epsilon)$ & & & $R$ & $:=$ &  $C \,|\, \epsilon$ \\
\end{tabular}
\end{center}
}

The C-O diagram \textbf{semantics} is defined by using NTAs ({\it Network of
  Timed Automata})
\cite{Alur94} as semantic objects. Here we omit this formal translation
and the technical definitions can be found in \cite{MCDS2012}. Instead, we
present an informal interpretation of the NTA behaviors.
When
transforming a C-O diagram into a network of timed automata, the nodes
of the generated automata
are decorated with the set of contractual obligations, prohibitions
and permissions that are either violated or satisfied.

\begin{definition}\label{def:violation-sets} (Violation, Satisfaction and Permissions Sets) 
  Let us consider the set of contractual obligations and prohibitions
  $CN$ ranged over $cn$, $cn'$,\ldots\
  standing for identifiers of obligations and prohibitions
  and the set of contractual permissions $CP$ ranged over $cp$, $cp'$,
  \ldots.
\end{definition}

\begin{definition}\label{def:decorated-automaton} (Decorated
  timed automaton)\\
  A decorated timed automaton is a timed automaton $(N, n_0, E, I)$
  (see~\cite{Alur94}) where for each $n\in N$ we have defined the following
  sets $V(n)\subseteq CN$ (the set of the obligations \emph{violated}
  in $n$), $S(n)\subseteq CN$ (the set of the obligations \emph{satisfied}
  in $n$), and $P(n)\subseteq CP$ (the set of permissions granted in $n$).
\end{definition}


Graphically, when we draw a timed automaton extended with these three
sets, we write under each node $n$ (between braces) its violation set
$V(n)$ on the left, its satisfaction set $S(n)$ on the centre and its
permission set $P(n)$ on the right. These sets are initially empty,
and they do not change except in two cases, a) when either a
obligation or a prohibition is violated or satisfied, b) when a
permission is performed.

Let us recall that the
intuitive meaning of an NTA is the parallel composition of several timed
automata.  We consider a set of actions $ACT$, in which we have the following actions:

\begin{multicols}{2}

\begin{itemize}
\item An internal action $\tau\in\act$.
\item An input action $m?\in\act$.
\item An output action $m!\in\act$.
\item A synchronization action $m\in\act$ that comes from a
  synchronization of an input action $m?$ and an output action $m!$.
\end{itemize}

\end{multicols}


The semantics of timed automata is well known \cite{Alur94}. It is based on a timed
labelled system, where states are pairs $s=(n,v)$ where $n$ is a node of the automaton
and $v$ is a valuation of the clocks. There are two types of transition:

\begin{multicols}{2}

\begin{itemize}
\item timed transitions\footnote{Timed transitions only change the valuation of
clocks.} $s\tran{d}s' (d \in \real)$
\item and action transitions $s\tran{a}s' (a \in ACT)$.
\end{itemize}

\end{multicols}


A {\it Network of Timed Automata}\, (NTA) is then defined as
a set of timed automata that
run simultaneously, using the same set of clocks,
and synchronizing on the common actions.
Internal actions can be executed by the corresponding automata
independently, and they will be ranged over the letters $a,b,\ldots$
whereas synchronization actions must be
executed simultaneously by two automata.
Synchronization actions are ranged over letters $m, m',\ldots$ and
they come from the synchronization of two
actions $m!$ and $m?$, executed from two different automata\footnote{In the
original definition the only internal action is $\tau$,
and synchronizations always yield internal actions.}.

The operational semantics of a network of timed automata has the
following transitions:


\begin{itemize}
\item A delay transition of $d$ time units requires that all the
  involved automata are able to perform this delay individually.

\item  Autonomous action transitions that correspond to the evolution
  of a single timed automaton.

\item Synchronization transitions that require two automata to perform
  two complementary actions, $m!$ and $m?$, respectively.

\end{itemize}

\begin{figure}

\begin{center}
  \includegraphics[width=12cm]{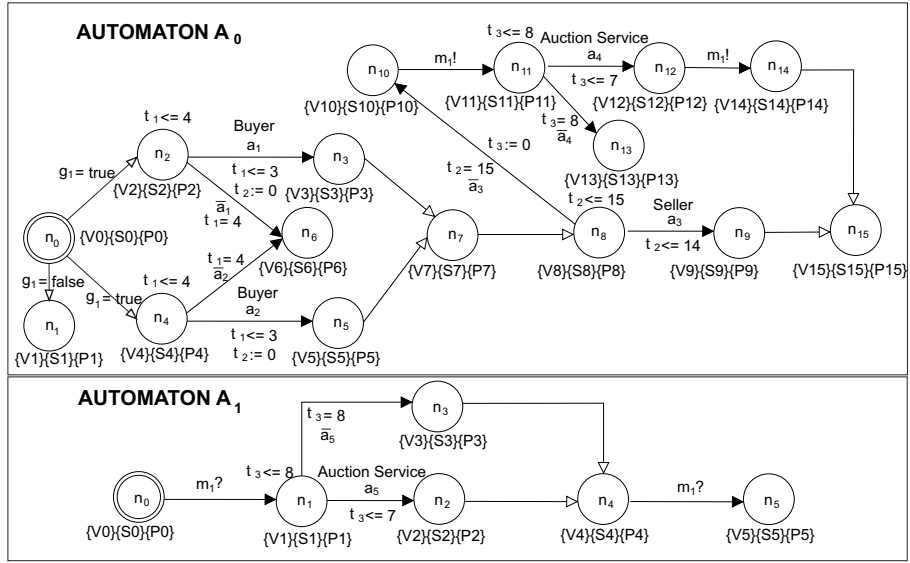}
\end{center}
  \vspace{-0.5cm}
  \caption{Automata for the Payment\_Shipment example, $A_0$ and $A_1$}
  \label{AutomatonOAP3}
\vspace{-0.5cm}
\end{figure}

\begin{definition}\label{def:nta-semantics} (Semantics of an NTA)\\
Let $N=(A_1, ,\ldots,A_k)$ be an NTA.
A \emph{state} of N is a tuple
$\overline{s}=(s_1,\ldots, s_k)$, where $s_i$  is
a state of the automaton $A_i$ (for $i=1,\ldots,k$).
We have the following transitions:

\begin{itemize}
\item Timed transitions. If \; $\forall 1\leq j\leq k:\ s_j\tran{d} s_j'$,
  then:
  $(s_1,\ldots, s_k)\tran{d} (s_1',\ldots, s_k')$ with $d \in \real$.
%
\item Autonomous transitions. If \; $\exists 1\leq j\leq k:\ s_j\tran{a}
  s_j'$ \; for $a\in\act$, $a \neq m!$ and $a\neq m?$, then:\\
  \hspace*{2.3cm} $(s_1,\ldots, s_j,\ldots s_k)\tran{a} (s_1,\ldots, s_j',\ldots s_k)$.

\item Synchronization transitions.
   $\exists 1\leq i,j\leq k:\ s_j\tran{m?} s_j',\ s_i\tran{m!} s_i'$ \;
   for $m?,m!\in\act$, then: \\
  $(\ldots, s_j,\ldots, s_i, \ldots,)\tran{m} (\ldots, s_j',\ldots, s_i', \ldots)$,
  assuming that $j \leq i$, the other case is similar.
\end{itemize}

\end{definition}

The complete semantics for \codiag in terms of NTAs translation can be found in \cite{MCDS2012}.
Figure \ref{AutomatonOAP3} shows the
resulting NTA once these transformations are applied over the \textit{Payment\_Shipment} example.
This NTA consists of two automata running
in parallel, that is, $NTA_{P\&S}=\{A_0,A_1\}$. Automaton $A_0$ is where the
main part is translated and the starting point of this example. The main
translated structures we can observe here are the three kind
of refinements and the reparation of a violated clause. Besides these
main structures, we can see how guards and time restrictions are translated.

In $A_0$, this contract starts with a SEQ-refinement of two clauses 2 and 5,
which assemble in sequence via the transition between nodes $n_7$ and $n_8$,
that is, the end of clause 2 and the beginning of clause 5, respectively.
From node $n_0$, where clause 2 starts, we may reach either $n_2$ or $n_4$,
which correspond to an OR-refinement representing the
payment made either by credit card or paypal. Node $n_6$
only captures termination in the event that that the time for the payment
expires without performing any of these actions. Notice that once the
payment has been done (nodes $n_3$ or $n_5$) we move into node $n_7$,
from which the ``sending item action'' clause 5 starts, which corresponds to
action $a_3$. In this case we have 14 time units. If this time expires
and the client has not received the item the reparation clause is
activated (node $n_{10}$). In this case we have an AND-refinement,
so a second timed automaton ($A_1$) is created, which corresponds to
the right-hand side part of the AND-refinement (the left-hand side is
performed by $A_0$). Both automata synchronize at their beginning and
at their termination in order to be
executed simultaneously. The obligation to refund the money is captured
by action $a_4$ in $A_0$, whereas the permission to penalize the seller is
captured by action $a_5$ in $A_1$. Over-line actions label those transitions
enabled when the main action is not performed.

Guards are here translated as guards in the transitions and time restrictions
are used to denote the invariants of certain states and some guards in transitions,
which determine whether a clause is satisfied in time. In reference to the different
violation, satisfaction and permission sets, we will comment the most significant ones,
which correspond with the maximal paths except when a reparation is defined.
Violation sets $V6$, $V10$ and $V13$ consist
of the violated clauses: either clause 3 or 4, clause 5 and clause 7, respectively.
The satisfaction set $S15$ will consist of clauses 3 or 4 (depending on the payment
is made either by credit card or paypal), and either clause 5
(if the item has been sent on time) or clause 7 (if the clause 5 has been repaired).
Finally, permission set $P15$ is either empty or clause 8 (if we have followed the
reparation and the permission to penalize the seller has been performed).



\section{Conformance relations}

In this section we define a set of conformance relations
to establish whether an implementation of a contract conforms
to the contract we want to satisfy. We will consider a semantic
relation inspired in the {\it conformance testing} relation
given in~\cite{tre99}. We take as starting point a normative document written in terms of
a C-O Diagram, which is then translated into a network of timed
automata. We also consider an implementation I of this contract
which is also provided as an NTA, with at least the same actions
we had in the contract.  We intend to define a black box conformance
relation, which means that we do not know how the implementation has
been done, so we can only use the information about the actions
it performs.

\begin{definition}
  A  \emph{timed trace} is a sequence $[a_1d_1 a_2d_2\cdots a_nd_n]\in (\act\times \real)^*$.
  We will use the symbols $t$, $t_1$, $t_2$, $t_n$,... to denote traces.
  The empty trace is denoted by $[]$. The concatenation of $t_1$ and $t_2$ will be
  denoted by $t_1\cdot t_2$. We will say that $t_1$ is a subtrace of $t_2$, written $t_1\leq t_2$,
  if there is a trace $t$ such that $t_2=t_1\cdot t$.

  Let $N$ be an NTA, where
  we define the  \emph{timed computations of $N$} as follows:
  \begin{itemize}
  \item $\overline{s}\vtran{[]} \overline{s}$.
  \item $\overline{s}\vtran{t\cdot [ad]} \overline{s'}$ for $a\in\act$ and $d\in\real$ if
    there exist states $\overline{s_1},\overline{s_1'},\ldots,
    ,\overline{s_l},\overline{s_l'}$ of $N$  with
    $l\geq 1$ such that\\
    $\overline{s}\vtran{t}\overline{s_1}\tran{d_1}\overline{s_1'}\tran{\tau}\overline{s_2}\tran{d_2}\overline{s_2'}\cdots\
   {\overline{s_{l-1}}} \tran{d_{l-1}} {\overline{s'_{l-1}}} \tran{\tau}{\overline{s_l}}\tran{d_l}\overline{s_l'}\tran{a}\overline{s'}$
    and $d=\sum_{1\leq i\leq l} d_i$
  \end{itemize}
  We define the \emph{set of timed traces} of $N$ as
  $\traces(N)=\{t\ |\ \exists \overline{s}:\ \overline{s_0}\vtran{t} \overline{s}\}$, being $\overline{s_0}$ the initial state of $N$.
\end{definition}


The following definition extends the sets V, S and P to traces, by accumulating
the contents of the respective sets V, S, P over the traversed nodes until
reaching the final node of the trace.

\begin{definition}
  Let $N=(A_1,\ldots A_k)$ be an NTA and $t \in \traces(N)$, we define the sets of violation (denoted $\tviolation(N,t)$),
  satisfaction (denoted $\tsatis(N,t)$), and permission (denoted $\tpermission(N,t)$) as follows:
  \begin{itemize}
  \item $\tviolation(N,t) = \{ \bigcup_{1\leq i\leq k} V(n_i)\ |\ \overline{s_0}\vtran{t} (s'_1,\ldots, s'_k),\ s'_i=(n_i, v_i)\}$
  \item $\tsatis(N,t) = \{ \bigcup_{1\leq i\leq k} S(n_i)\ |\ \overline{s_0}\vtran{t} (s'_1,\ldots, s'_k),\ s'_i=(n_i, v_i)\}$
  \item $\tpermission(N,t) = \{ \bigcup_{1\leq i\leq k} P(n_i)\ |\ \overline{s_0}\vtran{t} (s'_1,\ldots, s'_k),\ s'_i=(n_i, v_i)\}$
  \end{itemize}
  Where $\overline{s_0}$ is the initial state of $N$.
  We say that $t$ is a \emph{good} trace, denoted by $t\in\good(N)$ if it is maximal\footnotemark,
  $\forall S\in\tsatis(N,t):\ S\neq\emptyset$, and $\forall V\in\tviolation(N,t):\ V=\emptyset$.

  We say that $t$ is a \emph{clean} trace, denoted by $t\in\clean(N)$,
  if \; $\forall t'\leq t:\ \tviolation(N,t')=\{\emptyset\}$.
\end{definition}
\footnotetext{A maximal trace is a trace that cannot
  be extended anymore: if $t\in\traces(N)$ but $t\cdot
  [ad]\not\in\traces(N)$ for all $a\in\act$ and $d\in\real$.}

\begin{table}
\begin{center}
\begin{tabular}{|l||p{6cm}|c|c|c|c|}
\hline
Trace & Description & Nodes &  $V$ & $S$ & $P$ \\
\hline
\hline

$t_0=[\overline{a_1} 4]$ &  4 days without paying. & $(n_6,n_0)$ & 2 & $\emptyset$ & $\emptyset$ \\

$t_1=[a_1 3 a_3 8]$ & Credit card payment in 3 days and then item shipped in 8 days.  & $(n_{15},n_0)$ & $\emptyset$ & 3, 5 & $\emptyset$ \\

$t_2=[a_1 3 \overline{a_3} 15 \overline{a_4} 8]$ & Similar to $t_1$ but the item is not shipped. & $(n_{13},n_4)$ & 5 & 3 & $\emptyset$ \\

$t_3=[a_1 3 \overline{a_3} 15 a_5 2 \overline{a_4} 6]$ & Similar to $t_2$ but with a penalization. & $(n_{13},n_4)$ & 5 & 3 & 8 \\

$t_4=[a_2 2 \overline{a_3} 15 a_4 4]$ & Paypal payment in 2 days, item not received but refunded in 19 days. & $(n_{15},n_5)$ & $\emptyset$ & 4, 7, 5 & $\emptyset$ \\

$t_5=[a_2 2 \overline{a_3} 15 a_4 4 a_5 1]$ & Similar to $t_4$ but a penalization is made. & $(n_{15},n_5)$ & $\emptyset$ & 4, 7, 5 & 8 \\
\hline
\end{tabular}
\caption{Trace examples for $NTA_{P\&S}$.}
\label{traces}
\end{center}
\vspace{-0.5cm}
\end{table}

Comming back to our running example $NTA_{P\&S}$, let us analyse the following maximal traces of Table \ref{traces}.
The \emph{good} traces will be $t_1$, $t_4$ and $t_5$ since their violation sets are empty but
not their satisfaction sets. From these traces only $t_1$ corresponds to a \emph{clean} trace
since $t_4$ and $t_5$ have violated the shipment clause 5, however they have been recovered via $R_1$.


\begin{definition}
  Let $C$ be an NTA corresponding to a C-O diagram. We say that $C$ is consistent if the following
  conditions hold:
  \begin{itemize}
  \item $\clean(C)\cap\good(C)\neq\emptyset$.  This means that there is a way to meet contracts without
    making any violations.
  \item $\forall cn\in CN\ \exists t\in\clean(C)\cap\good(C):\ \exists S\in\tsatis(C,t):\ cn\in S$.
    That is there is a way to meet all obligations and prohibitions without making any violation.
  \end{itemize}
\end{definition}


Our $NTA_{P\&S}$ example satisfies both conditions since trace $t_1$ is a \emph{good} and \emph{clean} trace
that meets both obligations, the payment and the shipment.

As we have indicated previously, we assume that implementations
are given as
networks of timed automata.
Implementations usually need to implement a single action by making several simple actions.
For instance let us suppose that a contract specifies that a payment can be done by credit card. When
implementing the payment procedure, several invisible steps like connecting with the bank or checking
the credit card should be performed. All these actions are not considered in the specification of the contract
and they should not be taken into account. All we need in this case is
the amount of time required to
perform these actions. Thus, these implementation traces may contain actions that are not considered in the contract, so we
need to \emph{hide} these actions.

\begin{definition}
  Let us consider $\act\subseteq\act'$ and $t\in(\act'\times\real)^*$. We consider the operator
  $\hide{\act}$ defined as follows:
  \begin{itemize}
  \item $\hide{\act}([])=[]$
  \item $\hide{\act}([ad]\cdot t)=[ad]\cdot\hide{\act}(t)$ for $a\in\act$, $a\neq\tau$
  \item $\hide{\act}([ad]\cdot t)=d + \hide{\act}(t)$ for $a\not\in\act$
    or $a=\tau$,  where the operator $+$
    adds $d$ units of time to the last action of $t$. Formally it
    is defined as follows:
    \begin{itemize}
    \item $d+[]=[]$
    \item $d + ([ad_1] \cdot t)=[a(d_1+d)]\cdot t$
    \end{itemize}
  \end{itemize}
\end{definition}

Let us consider the following trace $t_6=[a_1 3 a_3' 2 a_3'' 2 a_3 4]$ belonging to a possible implementation
of our contract. The actions $a_3'$ and $a_3''$ are internal actions
of the implementation (for instance the seller
obtains the deliver company list related to the shipment address $a_3'$ and sends the shipment info to the deliverer $a_3''$).
Therefore, the result of $\hide{\act}(t_6)=[a_1 3 a_3 8]$,
where the internal actions have been omitted and the intermediate time
delays are 2 + 2 + 4 = 8.

Now, we have all the machinery needed to define our conformance relation. We will consider that
an implementation satisfies a contract if a) there is at least one trace
that execute all the actions expressed in the obligations in due time, and not any actions from the prohibitions; that
is, satisfying all the obligations and prohibitions expressed in the contract, and b) if at any time a violation
occurs, then it will be repaired in the future. In our example, the ideal implementation should be able to
 ``allow the user to at least pay with either credit card or paypal in 3 days, and then, the seller send the item in time''.
 This ideal behavior is represented by condition a), since it gathers all contract obligations and prohibitions.
 However we should be most realistic and think that all systems are prone to errors, then implementations can
 as well fail in some occasions. But if they do, then they should been able to recover somehow. That is the idea behind the second
 condition, that is, if a seller does not send the item, he should at least refund the buyer.

\begin{definition}
  Let us consider a consistent contract specification $C$ and an implementation $I$, we say that $I$ conforms $C$,
  written  $I\conf C$, iff
  \begin{itemize}
  \item For any $cn\in CN$ there exists $t\in\traces(I)$ such that $\hide{\act}(t)\in\clean(C)\cup\good(C)$ and
    $\exists S\in\tsatis(I,\hide{\act}(t)):\ cn\in S$.
  \item If there exists $t\in\traces(I)$ and $cn\in CN$ with
    $\exists V\in\tviolation(C,\hide{\act}(t)):\ cn\in V$, there exists $t'$ such that $t\cdot t'\in\traces(I)$
    such that $\hide{\act}(t\cdot t')\in\traces(C)$ and
    $\forall V'\in\tviolation(C,\hide{\act}(t\cdot t')):\ cn\not\in V'$.
  \end{itemize}
\end{definition}

Let us consider the following implementations $I_1$, $I_2$ and $I_3$ where $\traces(I_1)=\{t_1,t_2\}$, $\traces(I_2)=\{t_4\}$ and
$\traces(I_3)=\{t_1,t_4\}$ of our running example $NTA_{P\&S}$. The implementation $I_1$ satisfies the first condition since
$t_1$ is good and clean and satisfies all the $cn \in CN$, although it does not satisfies the second because $t_2$ violates clause 5,
 which is never repaired. Thus implementation $I_1$ does not conform the given contract. Regarding to
 $I_2$, we have the opposite situation, here trace $t_4$ violates the clause 5, but reparation $R_1$ is now applied to refund the buyer.
 Therefore this trace satisfies the second condition but not the first one because it does not includes all the $cn \in CN$. Finally, implementation $I_3$ is the only one that conforms the contract
 written as $I_3 \conf NTA_{P\&S}$, since it includes $t_1$ and $t_2$, which fulfil both conditions.



We are now interested in a comparison of different
implementations of a consistent contract, taking into
account the permissions allowed for each implementation.
This comparison will be based on the
permissions performed by an implementation in such a way that an implementation will
be considered \emph{better} than other if it is able to perform more permissions.
In our example we can consider two implementations, one that after the seller refunds
the buyer (because the item has not been sent), allows him to penalize the
seller; and other implementation, which does not allow penalizations. In this case,
we will say that the first one is better than the former one.

\begin{definition}
  Let us consider a consistent contract specification $C$ and two  implementations $I_1$ and
  $I_2$ such that $I_1\conf C$ and $I_2\conf C$. We say that $I_1$ is better with respect to
  the permissions than $I_2$, written $I_2\leqp I_1$  iff
  for any $t_2\in\traces(I_2)$ such that $\forall V\in\tviolation(C,\hide{\act}(t_2)): V=\emptyset$
  there is a trace $t_1\in\traces(I_1)$ such that $\forall V\in\tviolation(C,\hide{\act}(t_1)): V=\emptyset$
  and for any $P_1\in\permission(C,\hide{\act}(t_1))$ there exists
  $P_2\in\permission(C,\hide{\act}(t_2))$ such that $P_2\subseteq P_1$.
\end{definition}

Let us consider two new implementations $I_4$, $I_5$, where $\traces(I_4)=[t_1,t_4]$ and $\traces(I_5)=[t_1,t_5]$.
Both $I_4$ and $I_5$ conform to $C$, as they have at least
one trace ($t_1$) fulfilling all the obligations and prohibitions,
and traces $t_4$ for $I_4$ and $t_5$ for $I_5$ violate a clause,
but the corresponding reparation is performed on time.
That implies that both $I_4$ and $I_5$ $\conf$ $NTA_{P\&S}$.
However, the permission set for $I_4$ is empty,
whereas for $I_5$ clause 8 is the permission set. Thus,
$P_4 \subseteq P_5$, implying that $I_4 \leqp I_5$.

\begin{figure}
\begin{center}
\includegraphics[width=12cm]{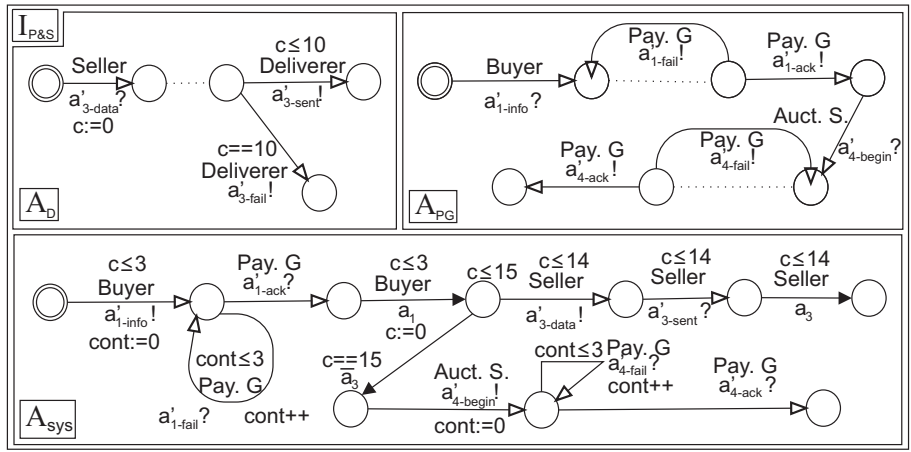}
\caption{Payment\_Shipment implementation example.}
\label{ImplementationExampleNTAP&S}
\end{center}
\end{figure}

In Figure \ref{ImplementationExampleNTAP&S}, an implementation
of the \emph{Paymen\_Shipment} example $I_{P\&S}$ is presented where
$I_{P\&S}=\{A_D,A_{PG},A_{SYS}\}$. $A_{SYS}$ is the main automata
where the main actions concerning to the contract are implemented.
$A_{D}$ and $A_{PG}$ implement the behaviors of a deliverer and
a payment gateway, both automata present some doted lines to describe
a set of internal actions that we abstract to simplify the example.
All the actions described in these two automata are synchronization actions
whose counterparts are defined in the main automata.
The deliverer automaton consists of three actions: the first one is
used to start the deliverer process by receiving the item data and
delivery address, then, the second one and the third are used to inform
the seller whether the delivery has succeed within
a time window of 10 days. The payment gateway is in charge of performing
two processes: charging the buyers credit card and perform the refund if
needed. They are performed via actions $a'_{1-...}$ and actions $a'_{4-...}$
where fail and ack actions are used to communicate whether the operation has
succeed or not, respectively.

Let us now analyze this implementation. The main answer to decipher is
if $I_{P\&S} \conf NTA_{P\&S}$. We can observe that the above defined
trace $t_1$ is obtained hiding the following trace
$t'_1=$$[a'_{1-info}0 $$a'_{1-ack} 3$ $ a_1 0$$ a'_{3-data} 0$ $ a'_{3-ack} 8$ $  a_3 0]$,
that is, $t_1=\hide{\act}(t'_1)$, and $t'_1 \in \traces(I_{P\&S})$. As we
have shown before, $t_1$ satisfies the first condition.
Regarding to the second condition, note that when a contract is broken it
is not necessary that the contract is always repaired, but it should exist
at least one trace allowing it\footnote{Cont variable is used to force the
refund for three times. If the refund is not feasible then a fail action is
executed.}. This occurs in trace
$t_4$, which can be obtained hiding the $I_{P\&S}$ trace
$t'_4=$ $[a'_{1-info}0 $ $a'_{1-ack} 2 $ $a_1 0$ $ \overline{a_3} 15 $ $a'_{1-begin} 0 $ $a'_{1-ack} 4 $ $a_4 0]$
and substituting
$a_2$ by $a_1$, that is, substituting the equivalent actions ``paypal'' payment
for a ``credit card'' payment. Thus, we show that the conformance relation is
held by our example, since it fulfils both criteria.

%
%

%
%
%
%



\section{Conclusions}
\label{Conclusions}

In this paper we have used the formal semantics based on NTAs
(Network of Timed Automata)
for normative contracts written in terms of C-O diagrams
introduced in \cite{MCD+10}
in order to define a conformance relation between a
contract and an implementation. We have
introduced the notion of {\em consistent} contracts on the
basis on their corresponding NTA,
as those NTAs that allow to find final traces without
violating any clauses. Then, implementations of
contracts are also NTAs, which must satisfy all the
obligations and prohibitions, or in the event of
a violation, implement the corresponding reparation.
These implementations are said to be conforming to the
contract. We have also presented a first comparison
relation between implementations, on the basis of
the permissions allowed for each one. We intend to
define a set of implementation comparisons,
taking into account the number of clauses that have
been violated, or assigning a weight to some clauses,
thus considering some clauses as more important.

\bibliographystyle{eptcs}
\bibliography{bibtex}

\begin{thebibliography}{10}
\providecommand{\bibitemdeclare}[2]{}
\providecommand{\surnamestart}{}
\providecommand{\surnameend}{}
\providecommand{\urlprefix}{Available at }
\providecommand{\url}[1]{\texttt{#1}}
\providecommand{\href}[2]{\texttt{#2}}
\providecommand{\urlalt}[2]{\href{#1}{#2}}
\providecommand{\doi}[1]{doi:\urlalt{http://dx.doi.org/#1}{#1}}
\providecommand{\bibinfo}[2]{#2}

\bibitemdeclare{article}{Alur94}
\bibitem{Alur94}
\bibinfo{author}{R.~\surnamestart Alur\surnameend} \& \bibinfo{author}{D.L.
  \surnamestart Dill\surnameend} (\bibinfo{year}{1994}):
  \emph{\bibinfo{title}{{A Theory of Timed Automata}}}.
\newblock {\sl \bibinfo{journal}{{Theoretical Computer Science}}}
  \bibinfo{volume}{{126(2)}}, pp. \bibinfo{pages}{183--235},
  \doi{10.1016/0304-3975(94)90010-8}.

\bibitemdeclare{article}{Chiu2003}
\bibitem{Chiu2003}
\bibinfo{author}{D.~\surnamestart Chiu\surnameend},
  \bibinfo{author}{S.~\surnamestart Cheung\surnameend} \&
  \bibinfo{author}{S.~\surnamestart Till\surnameend} (\bibinfo{year}{2003}):
  \emph{\bibinfo{title}{{A Three-Layer Architecture for E-Contract Enforcement
  in an E-Service Environment}}}.
\newblock {\sl \bibinfo{journal}{Proceedings of the 36th Hawaii International
  Conference on System Sciences (HICSS-36)}}, pp. \bibinfo{pages}{74--83},
  \doi{10.1109/HICSS.2003.1174188}.

\bibitemdeclare{article}{Dignum1995}
\bibitem{Dignum1995}
\bibinfo{author}{F.~\surnamestart Dignum\surnameend} \&
  \bibinfo{author}{H.~\surnamestart Weigand\surnameend} (\bibinfo{year}{1995}):
  \emph{\bibinfo{title}{{Modelling Communication between Cooperative
  Systems}}}.
\newblock {\sl \bibinfo{journal}{Proceedings of Advanced Information Systems
  Engineering (CAISE'95)}}, pp. \bibinfo{pages}{140--153},
  \doi{10.1007/3-540-59498-1_243}.

\bibitemdeclare{article}{Governatori2006}
\bibitem{Governatori2006}
\bibinfo{author}{G.~\surnamestart Governatori\surnameend},
  \bibinfo{author}{Z.~\surnamestart Milosevic\surnameend} \&
  \bibinfo{author}{S.~\surnamestart Sadiq\surnameend} (\bibinfo{year}{2006}):
  \emph{\bibinfo{title}{{Compliance checking between business processes and
  business contracts}}}.
\newblock {\sl \bibinfo{journal}{Proceedings of the 10th IEEE Conference on
  Enterprise Distributed Object Computing}}, pp. \bibinfo{pages}{221--232},
  \doi{10.1109/EDOC.2006.22}.

\bibitemdeclare{techreport}{Hatcliff2009}
\bibitem{Hatcliff2009}
\bibinfo{author}{J.~\surnamestart Hatcliff\surnameend}, \bibinfo{author}{G.T.
  \surnamestart Leavens\surnameend}, \bibinfo{author}{k.R.M. \surnamestart
  Leino\surnameend}, \bibinfo{author}{P.~\surnamestart Muller\surnameend} \&
  \bibinfo{author}{M.~\surnamestart Parkinson\surnameend}
  (\bibinfo{year}{2009}): \emph{\bibinfo{title}{{Behavioral Interface
  Specification Languages}}}.
\newblock \bibinfo{type}{Technical Report} \bibinfo{number}{CS-TR-09-01},
  \bibinfo{institution}{School of EECS, University of Central Florida},
  \doi{10.1145/2187671.2187678}.

\bibitemdeclare{article}{Krishna2005}
\bibitem{Krishna2005}
\bibinfo{author}{P.R. \surnamestart Krishna\surnameend},
  \bibinfo{author}{K.~\surnamestart Karlapalem\surnameend} \&
  \bibinfo{author}{A.R. \surnamestart Dani\surnameend} (\bibinfo{year}{2005}):
  \emph{\bibinfo{title}{{From Contract to E-Contracts: Modeling and
  Enactment}}}.
\newblock {\sl \bibinfo{journal}{Information Technology and Management}}
  \bibinfo{volume}{6}(\bibinfo{number}{4}), pp. \bibinfo{pages}{363--387},
  \doi{10.1007/s10799-005-3901-z}.

\bibitemdeclare{article}{Lomuscio2008}
\bibitem{Lomuscio2008}
\bibinfo{author}{A.~\surnamestart Lomuscio\surnameend},
  \bibinfo{author}{H.~\surnamestart Qu\surnameend} \&
  \bibinfo{author}{M.~\surnamestart Solanki\surnameend} (\bibinfo{year}{2008}):
  \emph{\bibinfo{title}{{Towards verifying contract regulated service
  composition}}}.
\newblock {\sl \bibinfo{journal}{Proceedings of IEEE International Conference
  on Web Services (ICWS 2008)}}, pp. \bibinfo{pages}{254--261},
  \doi{10.1109/ICWS.2008.115}.

\bibitemdeclare{article}{Marjanovic2001}
\bibitem{Marjanovic2001}
\bibinfo{author}{O.~\surnamestart Marjanovic\surnameend} \&
  \bibinfo{author}{Z.~\surnamestart Milosevic\surnameend}
  (\bibinfo{year}{2001}): \emph{\bibinfo{title}{{Towards formal modeling of
  e-Contracts}}}.
\newblock {\sl \bibinfo{journal}{Proceedings of 5th IEEE International
  Enterprise Distributed Object Computing Conference}}, pp.
  \bibinfo{pages}{59--68}, \doi{10.1109/EDOC.2001.950423}.

\bibitemdeclare{article}{Martinez2011}
\bibitem{Martinez2011}
\bibinfo{author}{E.~\surnamestart Mart\'{i}nez\surnameend},
  \bibinfo{author}{G.~\surnamestart D\'{i}az\surnameend} \&
  \bibinfo{author}{M.~E. \surnamestart Cambronero\surnameend}
  (\bibinfo{year}{2011}): \emph{\bibinfo{title}{{Contractually Compliant
  Service Compositions}}}.
\newblock {\sl \bibinfo{journal}{ICSOC 2011 - The Ninth International
  Conference on Service Oriented Computing}}, pp. \bibinfo{pages}{636--644},
  \doi{10.1007/978-3-642-25535-9_50}.

\bibitemdeclare{inproceedings}{MCD+10}
\bibitem{MCD+10}
\bibinfo{author}{E.~\surnamestart Mart\'{i}nez\surnameend},
  \bibinfo{author}{G.~\surnamestart D\'{i}az\surnameend},
  \bibinfo{author}{M.~E. \surnamestart Cambronero\surnameend} \&
  \bibinfo{author}{G.~\surnamestart Schneider\surnameend}
  (\bibinfo{year}{2010}): \emph{\bibinfo{title}{{A Model for Visual
  Specification of e-Contracts}}}.
\newblock In: {\sl \bibinfo{booktitle}{The 7th IEEE International Conference on
  Services Computing (IEEE SCC'10)}}, pp. \bibinfo{pages}{1--8},
  \doi{10.1109/SCC.2010.32}.

\bibitemdeclare{misc}{MCDS2012}
\bibitem{MCDS2012}
\bibinfo{author}{E.~\surnamestart Mart\'{i}nez\surnameend},
  \bibinfo{author}{G.~\surnamestart D\'{i}az\surnameend},
  \bibinfo{author}{M.~E. \surnamestart Cambronero\surnameend} \&
  \bibinfo{author}{G.~\surnamestart Schneider\surnameend}
  (\bibinfo{year}{2012}): \emph{\bibinfo{title}{{ Specification and
  Verification of Normative Specifications using C-O Diagrams}}}.
\newblock
  \bibinfo{howpublished}{\url{https://www.dsi.uclm.es/descargas/thecnicalreports/DIAB-12-05-1/TSE11.pdf}}.

\bibitemdeclare{incollection}{McNamara2006}
\bibitem{McNamara2006}
\bibinfo{author}{P.~\surnamestart McNamara\surnameend} (\bibinfo{year}{2006}):
  \emph{\bibinfo{title}{{Deontic Logic}}}.
\newblock In: {\sl \bibinfo{booktitle}{{Gabbay, D.M., Woods, J., eds.: Handbook
  of the History of Logic}}}, \bibinfo{volume}{7},
  \bibinfo{publisher}{North-Holland Publishing}, pp. \bibinfo{pages}{197--289},
  \doi{10.1016/S1874-5857(06)80029-4}.

\bibitemdeclare{inproceedings}{OR08csc}
\bibitem{OR08csc}
\bibinfo{author}{J.~C. \surnamestart Okika\surnameend} \&
  \bibinfo{author}{A.~P. \surnamestart Ravn\surnameend} (\bibinfo{year}{2008}):
  \emph{\bibinfo{title}{Classification of SOA Contract Specification
  Languages}}.
\newblock In: {\sl \bibinfo{booktitle}{2008 IEEE International Conference on
  Web Services (ICWS'08)}}, \bibinfo{publisher}{IEEE Computer Society}, pp.
  \bibinfo{pages}{433--440}, \doi{10.1109/ICWS.2008.36}.

\bibitemdeclare{inproceedings}{tre99}
\bibitem{tre99}
\bibinfo{author}{J.~\surnamestart Tretmans\surnameend} (\bibinfo{year}{1999}):
  \emph{\bibinfo{title}{Testing Concurrent Systems: A Formal Approach}}.
\newblock In: {\sl \bibinfo{booktitle}{CONCUR'99, LNCS 1664}},
  \bibinfo{publisher}{Springer}, pp. \bibinfo{pages}{46--65},
  \doi{10.1007/3-540-48320-9_6}.

\end{thebibliography}

\end{document}